\documentclass{article}

\usepackage{graphicx}
\usepackage{amsmath}
\usepackage{algorithm}
\usepackage{algorithmic}
\newcommand\eg{\emph{e.g.}}

\newcommand\bcmdtab{\noindent\bgroup\tabcolsep=0pt%
  \begin{tabular}{@{}p{10pc}@{}p{20pc}@{}}}
\newcommand\ecmdtab{\end{tabular}\egroup}

 \newcommand{\remove}[1]{}

\usepackage{color}
\newcommand{\changed}[1]{#1}

\title{The Impact of Network Flows on Community Formation in Models of Opinion Dynamics}

\author{Rumi Ghosh$^{1}$ and Kristina Lerman$^{2}$\\
	1. Robert Bosch LLC, Palo Alto, CA 94304, USA\\
         2. USC Information Sciences Institute, Marina del Rey, CA 90292, USA
         }

\begin{document}

\label{firstpage}

\maketitle

\begin{abstract}
We study dynamics of opinion formation in a network of coupled agents. As the network evolves to a steady state, opinions of agents within the same community converge faster than those of other agents. This framework allows us to study how network topology and network flow, \changed{which mediates the transfer of opinions between agents,} both affect the formation of communities.
In traditional models of opinion dynamics, agents are coupled via \changed{conservative flows, which result in one-to-one opinion transfer}. However, social interactions are often non-conservative, \changed{resulting in one-to-many transfer of opinions}.
We study opinion formation in networks using one-to-one and one-to-many interactions and show that they lead to different community structure within the same network.

\end{abstract}


\section{Introduction}
Networks often have complex structure that can be mapped onto communities in social networks or functional modules in biological networks~\cite{Ravasz2002Hierarchical,Rives03}. A community is generally understood to be a group of nodes that are better connected to, more similar to, or interact more frequently with, each other than with other nodes. Social scientists, mathematicians, physicists and computer scientists have developed an arsenal of methods for finding communities in networks.  These methods include clustering algorithms based on node similarity~\cite{Freeman01},  spectral clustering~\cite{tutorialspectral}, and graph partitioning methods based on conductance~\cite{Chung1997Spectral,Spielman:2004}, normalized cut~\cite{ShiMalik00}, and modularity~\cite{Fortunato2010Community}. What these methods have in common is that they examine network topology, or connections between nodes, to identify interesting structures.

Community structure of complex networks, however, is the product of both their topology \emph{and} function, which is determined by the dynamic processes taking place on the network. These processes, or \emph{flows}, mediate the interactions between nodes and determine the phenomena taking place on the network, whether diffusion and other types of transport in biological networks, or the spread of information or disease in social networks.
\changed{The relationship between network structure, topology and dynamics is complex and multi-faceted. Researchers have shown that topology and structure affect  the evolution of macroscopic phenomena taking place on a network, such as synchronization~\cite{Strogatz01,Boccaletti06,Pluchino06}, the spread of epidemics~\cite{PastorSatorras01} and rumors~\cite{Nekovee08}. The impact of structure is even greater when topology itself changes over time~\cite{Holme11}. A growing body of work examines the impact of dynamic interactions, or flows, on the measurement of network structure, including identification of central nodes~\cite{Borgatti05,Lambiotte10} and communities~\cite{Rosvall13} in the network. Our study of adds to this research by contrasting community structures induced by two different types of flow on the same network.
}

\changed{In this paper we study network community structure and its dependence on} both network topology and flows, using models of opinion dynamics. Such models have been used to explain how a social system can reach global consensus via local interactions. One class of models considers a network of coupled nodes, or agents, where each agent holds a real-valued opinion and affects \changed{the opinions of its neighbors by interacting with them}~\cite{socialDynamics,Pluchino06}.
In the long run, the network reaches a consensus state, with all agents holding the same opinion~\cite{Pluchino05}. However, strongly coupled agents --- those that interact more frequently or with greater intensity --- will converge in opinion faster. As a result, communities of agents holding similar opinions will emerge and coalesce  enroute to the global consensus state~\cite{Pluchino06}.

\changed{In existing models of opinion dynamics, agents transfer opinions to their neighbors}
via one-to-one, or \emph{conservative}, interactions.
Other examples of conservative interactions include used-goods and money exchange, in which an agent selects just one of its neighbors for a transaction. \changed{Such interactions are usually modeled as a random walk, whose dynamics is described by the Laplacian or one of its variants.} However, one-to-one interactions are not appropriate for describing other types social interactions, including those that lead to the spread of a disease and information. Such interactions are often {one-to-many}, or what we refer to as \emph{non-conservative}. In a simple epidemic, for example, an agent attempts to infect every neighbor, rather than pick a single neighbor to infect. Recently Lerman and Ghosh~\cite{Lerman12pre} introduced the replicator operator to describe the dynamics of agents coupled via non-conservative interactions. In Section~\ref{sec:synchronization} we \changed{use the replicator in a} model of opinion dynamics  and contrast its properties to those of the Laplacian.

\changed{As agents interact, either via conservative (one-to-one) or non-conservative (one-to-many) interactions, communities of agents holding similar opinions emerge in a complex network.} In Section~\ref{sec:communities} we define a function that measures the similarity of opinions of a pair of nodes or agents. We use this function to partition the network into communities of similar agents. In Section~\ref{sec:validation} we use the two interaction models to explore the community structure of real-world social networks of Facebook and Digg. We show that the non-conservative model identifies a different community structure than one that emerges from {conservative} interactions. While both models reveal a layered ``core and whiskers''~\cite{Leskovec08www} organization in the networks --- with a giant core and multiple small communities (whiskers) weakly connected to the core in each layer --- the composition of the cores found by the two models is substantially different. Moreover, the two models group into small communities subsets of nodes with different properties.

Our work \changed{highlights the fact that network structure, topology and dynamics are tightly interconnected. In order to identify meaningful structure, in addition to topology, community detection algorithms have to take into account the nature of interactions between network nodes.}

\section{Related Work}
Community detection is an extremely active research area, with a variety of methods proposed, including those based on similarity clustering~\cite{Freeman01}, spectral clustering~\cite{tutorialspectral} and graph partitioning methods that identify which edges to cut so as to  minimize conductance~\cite{Chung1997Spectral,Spielman:2004} or normalized cut~\cite{ShiMalik00}, or maximize modularity~\cite{GirvanNewman,Fortunato2010Community}.
These methods have been used to reveal the structure of complex networks. Leskovec et al.~\cite{Leskovec08www} found `core and whiskers' structure of real-world networks using conductance minimization and argued that this method cannot reveal any further structure in the giant core. Song~\cite{Song} claimed that there exist self-repeating patterns in complex networks at all length scales.  Our results corroborate these claims and show a repeating `core and whiskers' pattern in online social networks at many different length scales.

This paper studies the impact of dynamic interactions, or  flows, on a network's community structure using models of opinion dynamics. Such models attempt to explain the evolution of opinions in a network of \remove{locally interacting} coupled agents who can affect the opinion of their neighbors through local interactions. While a variety of approaches exist (for a review, please see \cite{socialDynamics}), we focus on models that describe real-valued, rather than discrete, opinions. In one such model~\cite{Pluchino05,Pluchino06}, agents attempt to align their opinions with those of their neighbors. Over time, groups of agents with similar opinions will emerge, and eventually coalesce as the network as a whole reaches a global consensus in which all agents hold the same opinion. These groups reveal the underlying community structure in the network. This model of opinion dynamics is equivalent to distributed synchronization, a well-studied physical phenomenon~\cite{Strogatz-book}.
Synchronization was first observed in the 17th century when clocks hung on the same wall synchronized the swing of their pendulums. Another famous example occurs in a population of fireflies who have characteristic light flashing patterns to help males and females recognize each other. Some firefly species appear to synchronize their flashing patterns with their neighbors, leading them to flash in unison. The Kuramoto model offers a simple mathematical description of synchronization in this and other physical and biological systems~\cite{Kuramoto}. The model considers a network of coupled oscillators, in which the phase (which could be taken as the opinion) of each node is affected by the difference between its phase and the phases of neighbors. While the network as a whole eventually reaches a fully synchronized (or consensus) state in which the phases of all nodes are the same, it does so in stages, with nodes belonging to the same community synchronizing faster than nodes belonging to different communities~\cite{Arenas06,Arenas2008Synchronization}.
The conservative interaction model described in this paper is a linear version of the Kuramoto model.

\changed{Several} researchers have explicitly studied how flows impact the measurement of network structure. Borgatti~\cite{Borgatti05} proposed that node's centrality reflects its participation in the flow taking place on the network, with different flows leading to different notions of centrality. However, he did not directly address the relationship between flows and network's community structure, though according to his arguments centrality is tied to group cohesiveness in networks~\cite{Borgatti06}.
Lambiotte et al.~\cite{Lambiotte10} proposed an integrated representation of the structure and dynamics of a network by embedding dynamic flows into edge weights of the adjacency matrix. While their framework is general and flexible enough to model the flows studied in this paper, they did not use it to find and compare community structure identified by different flows.
Rosvall et al.~\cite{Rosvall13} showed that introducing memory into a random walk in order to avoid nodes the walker has visited in the past, induces a different community structure on a network than an ordinary random walk.
This paper \changed{builds on these works by} demonstrating that \changed{details of the microscopic dynamics governing} flows affect the composition of cohesive groups, or communities, discovered within real-world social networks.

\section{Network Flows and Interaction Models}
\label{sec:synchronization}
We consider a network of $N$ active nodes (\eg, agents or actors). The state or opinion of node $i$ at time $t$ is described by a variable $\theta_i(t)$, which can change due to interactions with neighbors. As a result, the collective state of the network as a whole will also evolve over time. We represent the network as a weighted, undirected graph with a weight matrix $\boldsymbol{W}$, with $W[i,j]=w_{ij}$ representing the weight of an edge, or coupling strength, between nodes $i$ and $j$. If $i$ and $j$ are not connected, then $W[i,j]=0$. Another useful quantity is the degree matrix $\boldsymbol{D}$, a diagonal matrix with $D[i,i]=\sum_j w_{ij}$ and $D[i,j]=0$ for $i \ne j$.

We focus on a class of models in which  $\theta_i=\theta_i(t)$ is real-valued, and changes due to interactions with others. A variety of rules to update $\theta_i(t+1)$ were previously explored~\cite{socialDynamics}. In one model, which has been shown to lead to a global consensus~\cite{Pluchino06}, agents interact via one-to-one interactions. Many types of social interactions fall in this category. As an example, consider a network of agents exchanging books. When one agent decides to give a book to another, she chooses one of her neighbors and sends her the book. Money exchange and Web surfing are other examples of one-to-one interactions.
We refer to such interactions as \emph{conservative}, since they obey the principle of detailed balance. During book exchange, for example, one agent's loss of a book is offset by another one's gain.

Conservative interactions, however, cannot describe many other social interactions,  including those that lead to the spread of disease or information. Such interactions are often one-to-many: rather than picking one neighbor to infect, a sick individual will attempt to infect all neighbors. Since they do not obey detailed balance, we call such interactions \emph{non-conservative}. Below we present a model of opinion dynamics based on non-conservative interactions and study its properties.

\subsection{Conservative Interaction Model}

In a simple model of opinion dynamics~\cite{Pluchino06}, the degree to which the opinion of a neighbor $j$ impacts the opinion of node $i$ depends on the strength of their interaction, which is  proportional to the edge weight $w_{ij}$, and difference of opinions.\footnote{The specific model considered in \protect\cite{Pluchino06} exponentially attenuated the impact of $j$s opinion when it was very different from $i$s opinion. We consider the linear version of this model, since can be mapped to the linearized Kuramoto model, and more easily compared to non-conservative model described later in this paper.} The opinion of node $i$ then evolves according to:
\begin{eqnarray}
\label{eq:conservative}
\frac{d\theta_i}{dt} & = & \sum_j w_{ij}(\theta_j - \theta_i) \\ \nonumber
& = & -d_i \theta_i + \sum_j w_{ij} \theta_j
\end{eqnarray}
Here  $d_i=\sum_j w_{ij}$ is the (weighted) degree of node $i$.
Equation~\ref{eq:conservative} reveals the conservative nature of interactions. The change in opinion of node $i$ depends on the balance between the amount of opinion \emph{transferred to} neighbors at time $t$ (term $- d_i \theta_i$) and the amount \emph{received from} them at that time (term $\sum_j w_{ij} \theta_j $).

We can rewrite Eq.~\ref{eq:conservative} in matrix form as:
\begin{equation}
\label{eq:1}
\frac{d \boldsymbol{\theta}}{dt}=- \boldsymbol{L}\boldsymbol{\theta},
\end{equation}
where $\boldsymbol{\theta}$ is the vector of opinions, and $\boldsymbol{L}=\boldsymbol{D}-\boldsymbol{W}$ is the graph Laplacian matrix. Note that this model is identical to the linearized version of the Kuramoto model of synchronization~\cite{Arenas2008Synchronization}.

\subsection{Non-conservative Interaction Model}
Interactions need not always be conservative. Imagine, instead, that an agent broadcasts its opinion to all its neighbors, for example, through mass advertising, or posting it publicly on a social media site. In this non-conservative process opinions are replicated on each successful transmission. We define non-conservative interactions as follows. As before, each agent $i$ receives information from its neighbors ($\sum_j w_{ij} \theta_j$), but it does not transfer any to its neighbors and only loses it through a decay process at a rate $\alpha$. In this case, agent's opinion changes according to:
\begin{equation}
\label{eq:nonconservative}
\frac{d\theta_i}{dt} =  -\alpha \theta_i + \sum_j w_{ij} \theta_j
\end{equation}
Dynamics of the network can be written in matrix form as:
\begin{equation}
\label{eq:2}
\frac{d \boldsymbol{\theta}}{dt}= -(\alpha \boldsymbol{I}-\boldsymbol{W})\boldsymbol{\theta},
\end{equation}
where matrix $\boldsymbol{I}$ is the identity matrix.

The model above uses the \emph{replicator} operator $\boldsymbol{R}=(\alpha \boldsymbol{I}-\boldsymbol{W})$, the non-conservative counterpart of the graph Laplacian matrix~\cite{Lerman12pre}.
This operator governs the opinion dynamics of a network of agents coupled via non-conservative interactions.
In spite of non-conservation,
\changed{ a steady state exists in which opinions no longer change.}

\subsection{Steady State}
\changed{The solution to the conservative (Eq.~\ref{eq:1}) and non-conservative (Eq.~\ref{eq:2}) opinion dynamics models is given by:
\begin{equation}
\boldsymbol{\theta}({t})=\boldsymbol{\theta}_0 e^{-\boldsymbol{\mathcal{L}}(\boldsymbol{W})t}
\label{eq:8}
 \end{equation}
\noindent where $\boldsymbol{\mathcal{L}}(\boldsymbol{W})$ is the graph Laplacian $\boldsymbol{L}$ for the conservative model or the replicator $\boldsymbol{R}$ for the non-conservative model, and the initial value of each agent's opinion is $\theta_0=\theta(t=0)$. We assume that matrix $\boldsymbol{\mathcal{L}}(\boldsymbol{W})$ is diagonalizable and can be written as an eigenvalue decomposition $\boldsymbol{\mathcal{L}}(\boldsymbol{W})= \sum_{i=1}^N{\mathcal{X}[.,i]\lambda_i \mathcal{X}^{-1}[i,.]}$, where the $i$th column of $\mathcal{X}$ is the $i$th eigenvector of $\boldsymbol{\mathcal{L}}(\boldsymbol{W})$ with eigenvalue $\lambda_i$.
Eq.~\ref{eq:8}  can be written as:
 \begin{eqnarray}
\boldsymbol{\theta}({t})&=&\theta_{0}e^{-\boldsymbol{\mathcal{L}}t} =  \sum_{i=1}^N\mathcal{X}[.,i]e^{-\lambda_i t}\mathcal{X}^{-1}[i,.]\theta_{0} \nonumber \\
&=&\sum_{i=1}^N\mathcal{X}[.,i]e^{-\lambda_i t}c_i
 \label{eq:9}
 \end{eqnarray}
\noindent Here $c_i=\mathcal{X}^{-1}[i,.]\theta_{0}$ is a constant.

A non-trivial steady state $\boldsymbol{\theta}(t \to \infty) \ne 0$ exists when the smallest eigenvalue of $\boldsymbol{\mathcal{L}}(\boldsymbol{W})$ is 0, i.e., $\lambda_1=0$.
Under this condition, Eq.~\ref{eq:9} as $t\to \infty$ reduces to $\theta^{s}= \mathcal{X}[.,1]c_1$, with constant $c_1$.

In the conservative model, the smallest eigenvalue of $\boldsymbol{L}$ is zero in a connected network and the remaining eigenvalues are positive. Using Eq.~\ref{eq:9}, it is possible to show that a non-trivial steady state $\theta(t \to \infty) \neq 0$ exists in which $\theta_i=\theta_j$ $\forall i,\ j$~\cite{Arenas2008Synchronization}. In other words,  agents reach a consensus.

The steady state for a non-conservative model only exists for $\alpha=\lambda_{max}$, where $\lambda_{max}$ is the largest eigenvalue of $\boldsymbol{W}$. In this case $\boldsymbol{\theta}(t \to \infty)=\boldsymbol{\theta}^{s}$ is proportional  to the eigenvector of $\boldsymbol{W}$ corresponding to $\lambda_{max}$, also known as eigenvector centrality~\cite{Bonacich01}. This implies that the network becomes fragmented, with nodes holding different opinions and not yielding to the influence of others.
For $\alpha > \lambda_{max}$ the steady state has a trivial solution $\theta_i^{s} \to 0, \ \forall i$. Conversely, for $\alpha < \lambda_{max}$, the largest eigenvector is negative and values of $\theta_i$ diverge in the long term.
}

\subsection{Spectral Properties}
\changed{The spectrum of the dynamical operator gives information about the temporal and topological scales of opinion dynamics.} In conservative opinion dynamics, the rate at which the system asymptotically relaxes to the consensus steady state is determined by
the smallest positive eigenvalue of the Laplacian matrix $\boldsymbol{L}$.  Thus, the time to reach full consensus is inversely proportional to the smallest positive eigenvalue of $\boldsymbol{L}$, and the gaps between its consecutive eigenvalues are related to the relative difference in synchronization times of opinions of agents in different components~\cite{Arenas06,Arenas2008Synchronization}.


The convergence time of the non-conservative interaction model depends on the smallest positive eigenvalue of $\boldsymbol{R}=\lambda_{max}\boldsymbol{I}-\boldsymbol{W}$, where $\lambda_{max}$ is the largest eigenvalue of $\boldsymbol{W}$. Moreover, the spectrum of $\boldsymbol{R}$ gives insights into the community structure of the network. When the network has $C$ disjoint communities, the weight matrix $\boldsymbol{W}$ has $C$ eigenvalues that are significantly greater than the remaining $N-C$ eigenvalues~\cite{Chauhan09}. Since the eigenvalue spectrum of $\boldsymbol{R}=\lambda_{max}\boldsymbol{I}-\boldsymbol{W}$ is related to the spectrum of $\boldsymbol{W}$, in a network with $C$ disjoint communities, $\boldsymbol{R}$ has one null eigenvalue and $C-1$ eigenvalues that are much closer to zero than the remaining $N-C$ eigenvalues.

\changed{
The relationship between the eigenvalue spectrum of the dynamical operator and topological scales of the network forms the basis for spectral partitioning. Traditional spectral partitioning methods divide the network into communities based on the values of the eigenvectors of the graph Laplacian  (or its normalized version)~\cite{tutorialspectral,Chung1997Spectral}. Similarly, a network can be partitioned into communities based on epidemic-like processes using the replicator operator~\cite{Smith13pre}.
}

\section{Network Flows and Community Structure}
\label{sec:communities}
While spectral analysis can illuminate aspects of network structure \changed{and partition the network into communities}, simulating opinion dynamics on a network offers a more computationally efficient method to discover its community structure.  As demonstrated by Arenas et al.~\cite{Arenas06}, nodes' opinions in a conservative model converge in stages, with smaller units synchronizing their opinions before larger units, etc., until the entire network reaches a consensus. These stages reveal the hierarchical community structure of the network.
In this section we define a similarity function for interacting nodes and describe an algorithm for clustering them into communities.

\subsection{Similarity Measure}
We quantify the similarity of opinions of agents $i$ and $j$  at time $t$ using a function $s_{ij}(t)$:
\begin{equation}
s_{ij}(t)=\cos\big(\theta_i(t) - \frac{\theta_i^{s}}{\theta_j^{s}}\theta_j({t})\big),
\label{eq:sim}
\end{equation}
\noindent where $\theta_i^{s}$ ($\theta_j^s$) is the opinion node $i$ ($j$) holds in the steady state. The cosine function returns a value between one and zero, depending on the value of its argument.
As nodes synchronize their opinions, their similarity grows. Nodes are maximally similar ($s_{ij}(t)=1$) in the steady state, when $\theta_i=\theta_i^s$ and  $\theta_j=\theta_j^s$. In this situation, the argument of the cosine function is zero, and its value is one.

The rationale for this particular form of the similarity function is that when nodes reach the steady state, further interactions should not change their opinions. The same similarity function also applies to the conservative case. In this case all nodes have the same value in the steady state; therefore, Eq.~\ref{eq:sim} reduces to  the order parameter $s_{ij}=\cos(\theta_i(t) - \theta_j({t}))$ used by Arenas et al.~\cite{Arenas06}.

\subsection{Multi-Scale Community Detection}
We simulate dynamics by letting the network evolve according to the rules of the interaction model from some initial configuration. Generally, we choose random values of $\boldsymbol{\theta}$ as the initial configuration. At any time $t<t^{s}$, we can find the structure of the evolving network by clustering nodes using the similarity function $s_{ij}(t)$. \changed{We run each simulation multiple times and average the results.}

\begin{algorithm}
\caption{ Find communities after $t$ iterations at similarity threshold $\mu$}
\label{alg:2}
\begin{algorithmic}
\STATE{\bf{Input}}\\
$Y$: number of simulations of the interaction model $\mathcal{I}$ \\
$t$: number of iterations after which a network's evolving structure is analyzed\\
$\bar \theta_i(t)= (\theta_i(t)[1], \theta_i(t)[2], \cdots, \theta_i(t)[Y])$: values of $\theta_i(t)$ from all simulations.\\
$\mu=$similarity threshold \\
$e(i,j)$=edge between nodes $i$ and $j$ \\
\STATE{\bf{Output}}\\
Communities \{$C$\} such that $\forall i \in V$ $\max_{j \in C}(s_{ij}(t)) \ge (1-\mu)$ in the interaction model $\mathcal{I}$.\\
\STATE  {\bf{Initialize }}\\
$S=E$\\
Assign each node $i$ to a separate community $C_i \in C$.\\
\REPEAT
\FOR {each $e(i,j) \in E$}
    \STATE  $s_{ij}(t)= \frac{1}{Y}\sum_{y=1}^{Y}cos\big(\theta_i({t})^{[y]} - \big(\frac{\theta_{i}^{s}}{\theta_{j}^{s}}\big)^{[y]}\theta_j({t})^{[y]}\big) $
    \STATE $S=S-\{e(i,j)\}$
    \IF {$s_{ij}(t) \ge (1-\mu)$}
       \STATE Merge $C_i$ and $C_j$
    \ENDIF
\ENDFOR
\UNTIL{$S=\phi $}
\end{algorithmic}
\end{algorithm}

We can use the similarity function within any clustering procedure, e.g., a hierarchical agglomerative clustering algorithm. However, many such algorithms are not computationally efficient and cannot be used on large real-world networks. To deal with this problem, we use a simple coarse-graining algorithms that clusters nodes if their similarity is above some threshold.
Algorithm~\ref{alg:2} describes the clustering procedure that takes similarity threshold $\mu$ as input, and at time $t$ finds all communities in the network, such that if $i \in C_i$, $\max_{j \in C_i}(s_{ij}(t))$  is more than or equal to   $1-\mu$. Since by construction, in Algorithm \ref{alg:2},  for every $i \in C_i$, there exists a $j \in C_i$ , $1-\mu \le s_{ij}(t) \le \max_{j \in C_i}(s_{ij}(t))$, therefore in all communities output by this algorithm, for  all nodes $i \in C_i$, similarity $\max_{j \in C_i}(s_{ij}(t))\ge(1- \mu)$. This algorithm has linear runtime,  $O(|E|)$, where $|E|$ is the number of edges. By changing $\mu$, we can change the number and size of clusters. As $\mu$ increases, a cluster fragments into sub-clusters and thus a hierarchical arrangement of the clusters can be found.
The set of communities  output by Algorithm \ref{alg:2} after $t$ iterations of the simulation, for a given $\mu$ is unique  and independent of the order in which edges $e(i,j) \in E$ are considered.

The decentralized nature of the interaction models allows each node $i$ to compute $\theta_i$ locally interacting with at most $d_i$ of its neighbors, which helps us to parallelize the computation process making it fast and scalable.   Due to the linear nature of the interaction models considered, \changed{opinion dynamics model can be rewritten} as  $\sum_{k=1}^{N}{d \boldsymbol{\theta}(k)}/{dt}=- \boldsymbol{\mathcal{L}}(\boldsymbol{W}) \boldsymbol{\theta}(k) |_{\bar \theta_0(k)}$ where $\bar{\boldsymbol{\theta}}_0(k)[k]=\boldsymbol{\theta}_0[k]$ and is  $\bar{\boldsymbol{\theta}}_0(k)[j]=0$ $\forall$ $j\neq k$, $\boldsymbol{\theta}_0$ being the initial starting vector in Eq.~\ref{eq:8}. Each of the $N$ terms of this model can be calculated independently increasing  parallelizability further.

\section{Community Structure of Complex Networks}
\label{sec:validation}
We study the structure of real-world networks of social media sites Digg and Facebook by simulating opinion dynamics on their social graphs. We let opinions evolve from a random initial configuration, in which the values of $\theta_i$ are drawn from a uniform random distribution $[-\pi,\pi]$.
We identify as communities groups of nodes (users) holding similar opinions in the simulation. We contrast communities discovered by the conservative (Eq.~\ref{eq:1}) interaction model with those discovered by the non-conservative (Eq.~\ref{eq:2}) model by examining features of users within communities.
{We ran \changed{$Y=100$} simulations of each interaction model with different initial configurations and use these as input to the structure detection algorithm described in the previous section.}
\changed{For community analysis of synthetic and benchmark networks using this method, please see \cite{Lerman12pre}.}

\subsection{Digg Mutual Follower Network}
{Digg} (http://digg.com) is a social news aggregator with over 3 million registered users.  Users submit links to news stories and recommend them to other users by voting on, or {digging}, them. Digg also allows users to follow other users to see the new stories they have recently submitted or voted for. We extracted data about all users who voted on stories featured on Digg's front page in June 2009. Our data also includes the follower graph of the voters.\footnote{http://www.isi.edu/$\sim$lerman/downloads/digg2009.html}
 From this data, we reconstructed undirected mutual follower network in which an edge between $a$ and $b$ means that user $a$ follows user $b$ \textbf{and} $b$ follows $a$. This network comprises of around 40K nodes and more than 360K edges.
There are 4,811 disconnected components, with the largest component containing 27K nodes and 352K edges. The second largest component has 22 nodes. Since the inherent richness of structure of this network is largely captured by the giant component, we study this component in detail.

 \begin{figure}[t]
\centering
\begin{tabular}{@{}c@{}c@{}}
\includegraphics[width=0.52\linewidth]{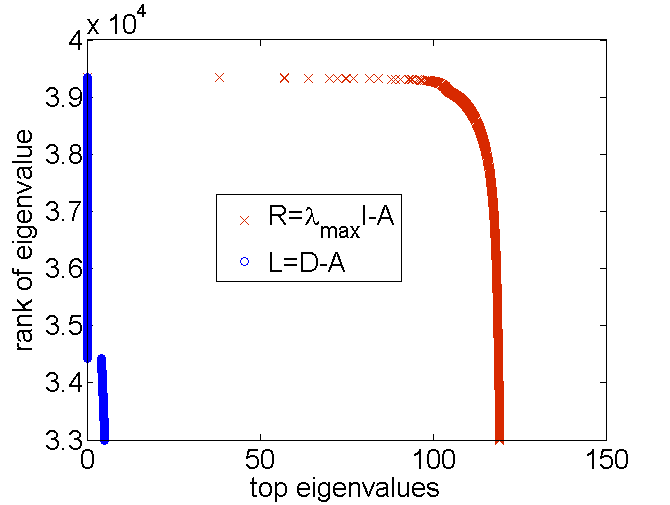} &
\includegraphics[width=0.48\linewidth]{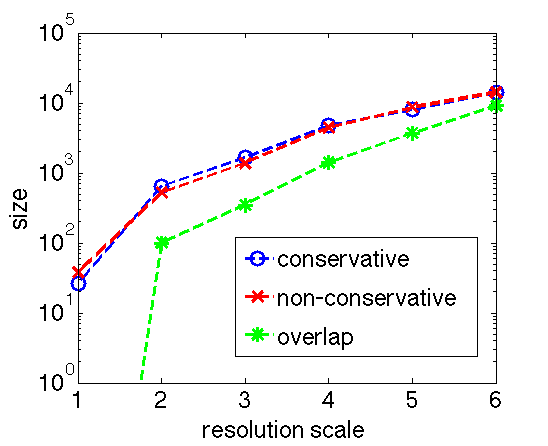} \\
(a) & (b)
\end{tabular}
\caption{(a) Top 6000 eigenvalues of the Replicator and Laplacian operators of the Digg friendship network. (b)
Number of nodes nodes comprising the core found by the interaction models at different resolution scales.
}
 \label{fig:diggcore}
\end{figure}

Using the Jacobi-Davidson Algorithm for calculating eigenvalues of a graph,  we compute more than 6K of the smallest eigenvalues of the Replicator and  Laplacian  operators and rank them in descending order (Fig.~\ref{fig:diggcore}(a)). The two spectra are dramatically different. The smallest positive eigenvalue of $\boldsymbol{L}$ is much smaller than that of $\boldsymbol{R}$. This indicates that the non-conservative interaction model reaches the steady state much faster than the conservative model.

\subsubsection{Multi-scale Structure of Digg}
We use Algorithm~\ref{alg:2} to cluster nodes at different resolution scales, specified by similarity threshold $\mu$. Both interaction models identify an intricate multi-scale organization of the network, though the overall structure may change over time.
Specifically, both models find a multi-layer `core and whiskers' structure~\cite{Leskovec08www} in the network, with one giant community (core) and many small communities (whiskers) loosely connected to the core. Upon changing the resolution scale (similarity threshold), the core breaks up into another core and many whiskers.
Thus, the community structure of Digg resembles an onion, with multiple layers of whiskers.  Figure~\ref{fig:diggcore}(b) shows sizes of cores at different resolution scales found by the two interaction models at time $t=100$.
For each model, we chose the threshold parameters $\mu$ that give comparable size cores. In the non-conservative model, all thresholds above $\mu=0.0004$ (resolution scale 7) produce a single component with 27K nodes.  At a finer resolution (decreasing $\mu$), we find another large core, whose size is reported in Fig.~\ref{fig:diggcore}(b), and many small communities. This trend continues, until $\mu=0.000008$ (resolution scale 1), when the core decomposes into several small communities.

While the onion-like organization discovered by both models is similar, its composition is very different. Figure~\ref{fig:diggcore}(b) reports the overlap of membership of comparable-size cores found by the two models. For example, the size of the giant component discovered by non-conservative interaction model for $\mu=0.00018$ is comparable to the size of the core discovered by the conservative interaction model for $\mu=0.2$; however, they share only about $80\%$ of the nodes.
Core overlap decreases to about $40\%$ at $\mu=0.00014$ for non-conservative interaction model ($\mu=0.008$ for conservative model), and keeps on decreasing as we fine-tune the resolution scale. Finally, the cores found at $\mu=0.00008$ for non-conservative and $\mu=0.0001$ for conservative models (resolution scale 1) do not have any nodes in common.

\subsubsection{Properties of Small Communities}

\begin{figure}[tbh]
\center
\begin{tabular}{@{}c@{}c@{}}
  \includegraphics[width=0.5\linewidth]{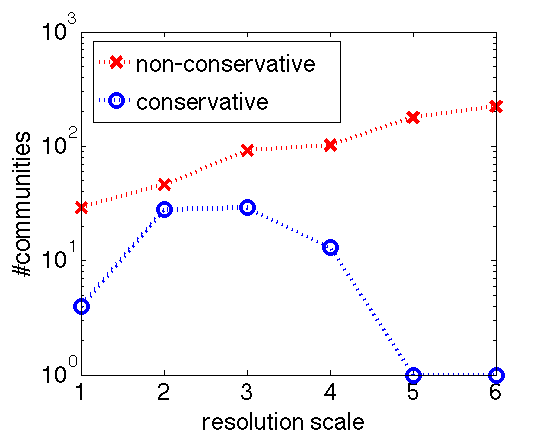} &
  \includegraphics[width=0.5\linewidth]{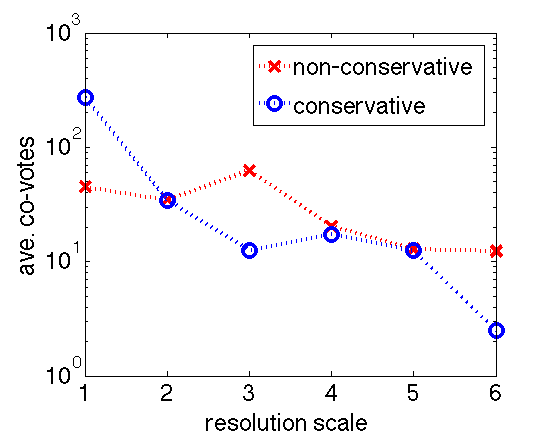}  \\
  (a) & (b)
\end{tabular}
  \caption{Properties of small communities found in the Digg mutual follower graph at $t=100$ by the two interaction models. (a) Number of small communities at different resolutions specified by the similarity threshold parameter. The smallest resolution scale corresponds to smallest value of similarity threshold. (b) Average number of co-votes made by community members.}\label{fig:digg_results}
\end{figure}

We now focus on small communities (whiskers) isolated from the core at different resolution scales. A whisker is a community of at least size three.  Figure~\ref{fig:digg_results}(a) shows the number of small communities resolved by the two interaction models at different scales.
Non-conservative interaction model assigned 3,712 distinct users to such communities (when summed over all resolution scales in Figure \ref{fig:digg_results}(a)). In contrast, the conservative interaction model assigned just 449 distinct users to small communities. The rest of the users fragmented into isolated pairs or singletons.

Besides their size and number, how do the nodes assigned to small communities by each interaction model differ?
We measure similarity of two Digg users by the number of  stories for which they both voted, i.e., \emph{co-votes}. Then, averaging over co-votes of all connected pairs of community members, we obtain a measure of community ``cohesiveness.''
As seen in Figure \ref{fig:digg_results}(b), average number of co-votes increases at finer resolution scales, producing more cohesive communities in the center of the `onion'. Members of the innermost communities (resolution scale 1), are much more similar according to the average number of co-votes than members of the outer communities (resolution scales 5, 6). Except at resolution scale 1, the average cohesiveness of communities found by the non-conservative model is higher than that found by the conservative model.  The difference at resolution scale 1 is driven by the two outliers in the conservative model. The first of these is a community of 26 users, with more than 300 co-votes on average, and the other is a community of nine with more than 600 co-votes. In addition to co-voting on an extraordinary number of stories (600 is nearly 20\% of all stories in our data set), these users are also highly interlinked. The first group forms a 13-core (a cluster in which each node is linked to at least 13 other nodes), and the second group forms a 4-core. These users also share many friends. While we cannot say whether these groups represent the often-rumored voting blocs on Digg, their activity does appear to be anomalous.
To summarize, non-conservative model finds many more small communities that are more cohesive than the conservative model, though the latter seems to pick out anomalous groups of users.

\begin{figure}[tbhp]
\center
\begin{tabular}{@{}c@{}c@{}}
  \includegraphics[width=0.5\linewidth]{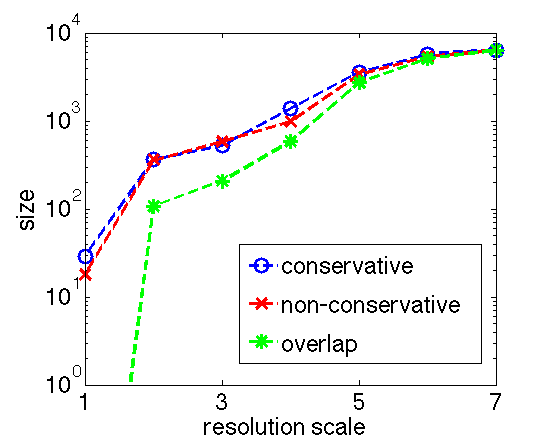} &
  \includegraphics[width=0.5\linewidth]{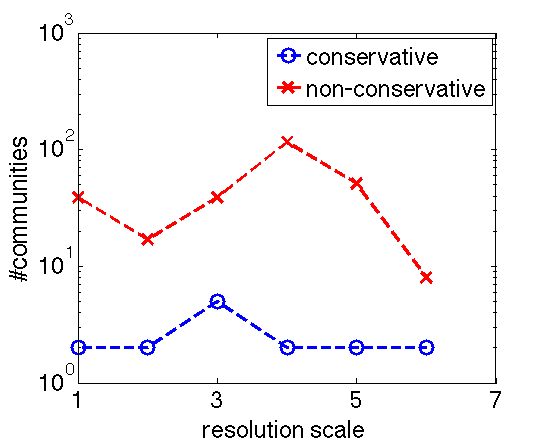}  \\
  (a) & (b)
\end{tabular}
  \caption{Distribution of communities in the Facebook network for American University at $t=100$. (a) Number of nodes nodes comprising the core found by the interaction models at different resolution scales. (b)  Number of small communities at different resolutions. The smallest resolution corresponds to highest similarity between individuals.}\label{fig:fb_comms}
\end{figure}

\subsection{Facebook Social Network}
We also analyzed a data set containing a snapshot of the Facebook networks as of September 2005~\cite{facebook}. Each user in this data set has four descriptive features: status (e.g., student, faculty, staff, and so on),  major, dorm or house, and graduation year. We use these features to measure similarity of members of the discovered communities. While this data set contains more than 100 colleges and universities, we present here the analysis of the network for American University, which comprises of 6,386 nodes and more than 200K edges.

\subsubsection{Multi-scale Structure of Facebook}
We use Algorithm~\ref{alg:2} to cluster nodes at different resolution scales specified by the similarity threshold $\mu$. As with Digg, we find a multi-scale organization in the structures discovered by conservative and non-conservative interaction models.
At each resolution scale, we find a giant core and many small communities. As with Digg, there is little overlap in membership between cores found by the two interaction models at finer resolutions (Fig.~\ref{fig:fb_comms}(a)).

As on Digg, many nodes participate in small, clique-like communities.  Figure~\ref{fig:fb_comms}(b) shows the number of communities discovered at each resolution scale by conservative and non-conservative models.
While 1,320 nodes contribute to the formation of such communities in the non-conservative model, only 32 nodes participate in such communities in the conservative interaction model (summed over resolution scales 1 to 7 in Figure \ref{fig:fb_comms} (b)). The remaining users are fragmented into isolated pairs or singletons. As in the Digg network, non-conservative model found many more communities than the conservative model.

\begin{figure}[tbh]
\center
\begin{tabular}{@{}c@{}c@{}}
    \includegraphics[width=0.5\linewidth]{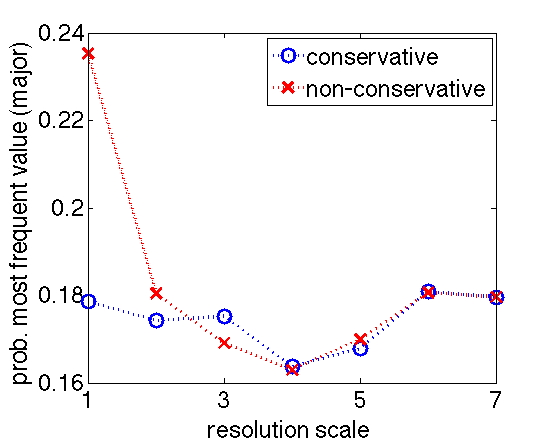} &
  \includegraphics[width=0.5\linewidth]{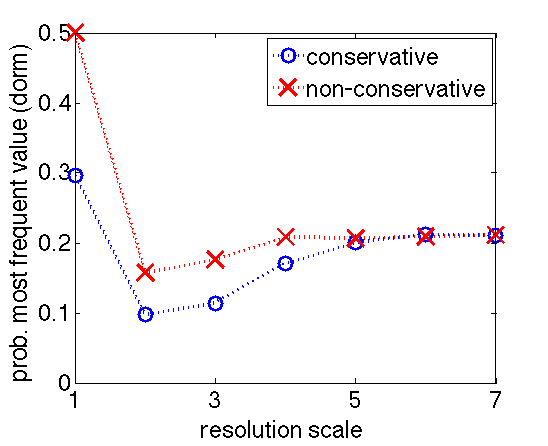}  \\
  (a) &  (b)\\
      \includegraphics[width=0.5\linewidth]{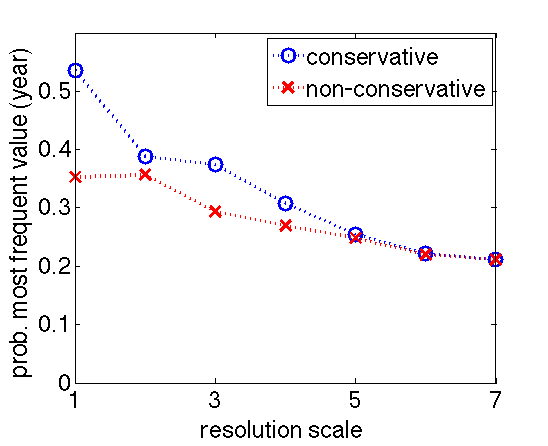} &
  \includegraphics[width=0.5\linewidth]{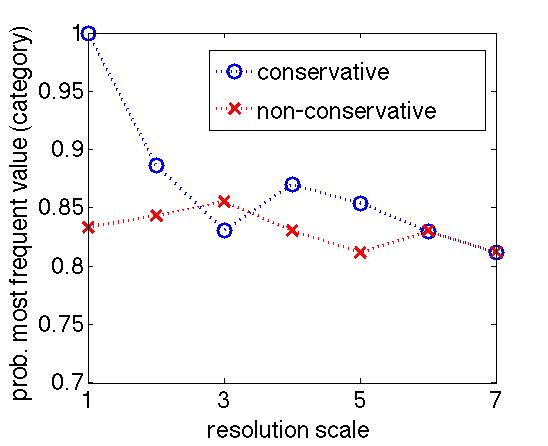}  \\
  (c) & (d)
\end{tabular}
  \caption{Properties of small communities found in the Facebook network of American University at $t=100$ by the two interaction models.  Each plot shows at different resolution scales the probability of occurrence of the most frequent value of features (a) major, (b) dorm, (c) year, (d) category of individual. }\label{fig:fb_results}
\end{figure}

\subsubsection{Properties of Small Communities}
How do the small communities discovered at different resolution scales by the two models in the Facebook network differ? We look at four features of users in the data set --- major, dorm, year and category of individual --- and calculate the prevalence of  feature values among community members. The community is characterized by the prevalence of the most popular feature among its members, or its cohesiveness with respect to that feature. For example, when using the dorm feature to characterize the community, dorm cohesiveness is the largest fraction of community members that belong to the same dorm.

Figure~\ref{fig:fb_results} shows the cohesiveness of communities found by the two models at different resolution scales with respect to these features. The prevalence of the dominant feature increases at smaller scales (tighter similarity threshold), irrespective of the feature under consideration. However, the characteristics of the community structure discovered by conservative and non-conservative interaction models vary significantly.
At finer resolution scales, non-conservative model finds communities of individuals who are more likely to have the same major and belong to the same dorm. Conservative model, on the other hand, is more likely to put into the same community individuals who belong to the student category and are in the same year.
These results suggest a possibility that students who belong to the same year may have more face to face (conservative) interactions, while students who have the same major or live in a dorm, may meet in study groups, or organized events and in the cafeteria, increasing chances for one-to-many (non-conservative) interactions.

In summary, regardless of the interaction process, we observe a roughly scale invariant organization in the real-world social networks. At  almost every resolution scale, we find a large component and many small components or communities.  Thus, Digg and Facebook's structure resembles an onion. Peeling each layer reveals another, almost self-similar structure with a core and many smaller communities. However, the exact composition of communities depends on the specifics of the interaction process.

\section{Conclusion}
Our work highlights the importance of network flows in the analysis of network structure and provides a framework for \changed{understanding how topology and dynamic flows jointly contribute to community detection}.
We investigated the interplay between structure and dynamics using models of opinion dynamics. As nodes within a network interact, their initially disparate opinions become more similar, with strongly coupled nodes within the same community synchronizing their opinions faster than other nodes. This observation allows us to use dynamics of opinion formation as a basis for finding communities in networks. We also proposed a class of opinion dynamics models based on non-conservative interactions and analyzed their properties, such as conditions for which a steady state exists.

Our study of the community structure of \changed{large-scale online} social networks revealed a complex `onion'-like organization. Peeling each level of hierarchy gives a core and many small components, regardless of the interaction model. However, different interactions lead to different views of this multi-scale organization, with conservative and non-conservative models uncovering communities that differ in size, composition, and properties of nodes.
Conservative and non-conservative processes represent just two types of flows in networks. It would be interesting to discover and mathematically characterize other types of network flow and determine their impact on community structure.

\subsection*{Acknowledgements}
This material is based upon work supported in part by  the Air Force Office of Scientific Research under Contract Nos. FA9550-10-1-0569,  by the Air Force Research Laboratories under contract FA8750-12-2-0186, by the National Science Foundation under Grant No. CIF-1217605, and by DARPA under Contract No. W911NF-12-1-0034.

\bibliographystyle{plain}
\bibliography{references}  

\begin{thebibliography}{10}

\bibitem{Arenas2008Synchronization}
Alex Arenas, Albert D\'{\i}az-Guilera, Jurgen Kurths, Yamir Moreno, and
  Changsong Zhou.
\newblock {Synchronization in complex networks}.
\newblock {\em Physics Reports}, 469(3):93--153, December 2008.

\bibitem{Arenas06}
Alex Arenas, Albert~D. Guilera, and Conrad~J. P\'{e}rez~Vicente.
\newblock {Synchronization Reveals Topological Scales in Complex Networks}.
\newblock {\em Physical Review Letters}, 96(11):114102+, March 2006.

\bibitem{Boccaletti06}
S.~Boccaletti, V.~Latora, Y.~Moreno, M.~Chavez, and D.~Hwang.
\newblock Complex networks: Structure and dynamics.
\newblock {\em Physics Reports}, 424(4-5):175--308, February 2006.

\bibitem{Bonacich01}
Phillip Bonacich and Paulette Lloyd.
\newblock Eigenvector-like measures of centrality for asymmetric relations.
\newblock {\em Social Networks}, 23(3):191--201, 2001.

\bibitem{Borgatti05}
S.~Borgatti.
\newblock Centrality and network flow.
\newblock {\em Social Networks}, 27(1):55--71, January 2005.

\bibitem{Borgatti06}
S.~Borgatti and M.~Everett.
\newblock A graph-theoretic perspective on centrality.
\newblock {\em Social Networks}, 28(4):466--484, October 2006.

\bibitem{socialDynamics}
Claudio Castellano, Santo Fortunato, and Vittorio Loreto.
\newblock Statistical physics of social dynamics.
\newblock {\em Reviews of Modern Physics}, 81(2):591--646, May 2009.

\bibitem{Chauhan09}
Sanjeev Chauhan, Michelle Girvan, and Edward Ott.
\newblock Spectral properties of networks with community structure.
\newblock {\em Physical Review E}, 80(5):056114+, November 2009.

\bibitem{Chung1997Spectral}
Fan R.~K. Chung.
\newblock {\em {Spectral Graph Theory (CBMS Regional Conference Series in
  Mathematics, No. 92)}}.
\newblock American Mathematical Society, February 1997.

\bibitem{Fortunato2010Community}
Santo Fortunato.
\newblock Community detection in graphs.
\newblock {\em Physics Reports}, 486:75--174, January 2010.

\bibitem{Freeman01}
Linton~C. Freeman.
\newblock Finding social groups: A meta-analysis of the southern women data.
\newblock In {\em Dynamic Social Network Modeling and Analysis. The National
  Academies}, pages 39--97. Press, 2003.

\bibitem{GirvanNewman}
M.~Girvan and M.~E.~J. Newman.
\newblock Community structure in social and biological networks.
\newblock {\em Proc. Natl. Acad. Sci. USA.}, 99:7821, 2002.

\bibitem{Holme11}
Petter Holme and Jari Saram\"{a}ki.
\newblock Temporal networks.
\newblock {\em Physics Reports}, 519(3):97--125, December 2011.

\bibitem{Kuramoto}
Y.~Kuramoto.
\newblock {\em Chemical Oscillations, Waves, and Turbulence}.
\newblock Dover, Mineola, NY, 2003.

\bibitem{Lambiotte10}
R.~Lambiotte, R.~Sinatra, J.~C. Delvenne, T.~S. Evans, M.~Barahona, and
  V.~Latora.
\newblock Flow graphs: interweaving dynamics and structure.
\newblock December 2010.

\bibitem{Lerman12pre}
Kristina Lerman and Rumi Ghosh.
\newblock Network structure, topology and dynamics in generalized models of
  synchronization.
\newblock {\em Physical Review E}, 86(026108), 2012.

\bibitem{Leskovec08www}
Jure Leskovec, Kevin~J. Lang, Anirban Dasgupta, and Michael~W. Mahoney.
\newblock {Statistical properties of community structure in large social and
  information networks}.
\newblock In {\em WWW}, pages 695--704, New York, NY, USA, 2008. ACM.

\bibitem{Nekovee08}
Maziar Nekovee, Y.~Moreno, G.~Bianconi, and M.~Marsili.
\newblock Theory of rumour spreading in complex social networks.
\newblock {\em Physica A: Statistical Mechanics and its Applications},
  374(1):457--470, July 2008.

\bibitem{Pluchino06}
Alessandro Pluchino, Stefano Boccaletti, Vito Latora, and Andrea Rapisarda.
\newblock Opinion dynamics and synchronization in a network of scientific
  collaborations.
\newblock {\em Physica A: Statistical Mechanics and its Applications},
  372(2):316--325, December 2006.

\bibitem{Pluchino05}
Alessandro Pluchino, Vito Latora, and Andrea Rapisarda.
\newblock Changing opinions in a changing world: A new perspective in
  sociophysics.
\newblock {\em International Journal of Modern Physics C}, 16(04):515--531,
  2005.

\bibitem{Ravasz2002Hierarchical}
E.~Ravasz, A.~L. Somera, D.~A. Mongru, Z.~N. Oltvai, and A.~L. Barab\'{a}si.
\newblock {Hierarchical Organization of Modularity in Metabolic Networks}.
\newblock {\em Science}, 297(5586):1551--1555, 2002.

\bibitem{Rives03}
Alexander~W. Rives and Timothy Galitski.
\newblock Modular organization of cellular networks.
\newblock {\em Proc Natl Acad Sci U S A}, 100(3):1128--1133, 2003.

\bibitem{Rosvall13}
Martin Rosvall, Alcides~V. Esquivel, Andrea Lancichinetti, Jevin~D. West, and
  Renaud Lambiotte.
\newblock Networks with memory, July 2013.

\bibitem{PastorSatorras01}
Romualdo~P. Satorras and Alessandro Vespignani.
\newblock Epidemic spreading in {Scale-Free} networks.
\newblock {\em Physical Review Letters}, 86(14):3200--3203, April 2001.

\bibitem{ShiMalik00}
J.~Shi and J.~Malik.
\newblock Normalized cuts and image segmentation.
\newblock {\em IEEE Transactions on Pattern Analysis and Machine Intelligence},
  22(8):888--905, 2000.

\bibitem{Smith13pre}
Laura~M. Smith, Kristina Lerman, Cristina Garcia-Cardona, Allon~G. Percus, and
  Rumi Ghosh.
\newblock Spectral clustering with epidemic diffusion.
\newblock {\em Physical Review E}, 88(4):042813, 2013.

\bibitem{Song}
Chaoming Song, Shlomo Havlin, and Hernan~A. Makse.
\newblock {Self-similarity of complex networks}.
\newblock {\em Nature}, 433(7024):392--395, January 2005.

\bibitem{Spielman:2004}
Daniel~A. Spielman and Shang-Hua Teng.
\newblock Nearly-linear time algorithms for graph partitioning, graph
  sparsification, and solving linear systems.
\newblock In {\em Proceedings of the thirty-sixth annual ACM symposium on
  Theory of computing}, STOC '04, pages 81--90, New York, NY, USA, 2004. ACM.

\bibitem{Strogatz-book}
Steven Strogatz.
\newblock {\em Sync: The Emerging Science of Spontaneous Order}.
\newblock Theia, March 2003.

\bibitem{Strogatz01}
Steven~H. Strogatz.
\newblock Exploring complex networks.
\newblock {\em Nature}, 410(6825):268--276, March 2001.

\bibitem{facebook}
Amanda~L. Traud, Eric~D. Kelsic, Peter~J. Mucha, and Mason~A. Porter.
\newblock Comparing community structure to characteristics in online collegiate
  social networks.
\newblock {\em SIAM Rev.}, 53(3):526--543, August 2011.

\bibitem{tutorialspectral}
Ulrike von Luxburg.
\newblock {A tutorial on spectral clustering}.
\newblock {\em Statistics and Computing}, 17(4):395--416, December 2007.

\end{thebibliography}


\end{document}